\documentclass[prl,twocolumn,floatfix,superscriptaddress]{revtex4}
\usepackage{amssymb,amsmath}
\usepackage{graphicx}
\usepackage{subfigure}

\newcommand{\mb}[1]{\ensuremath{\mathbf{#1}}}
\newcommand{\h}[1]{\ensuremath{\hat{\mb{#1}}}}

\begin{document}
\title{Simulations of strongly phase-separated liquid-gas systems}

\author{A.J. Wagner} 

\affiliation{Department of Physics, North Dakota State University,
Fargo, ND 58105}
\email{alexander.wagner@ndsu.edu}
\author{C.M. Pooley}
\affiliation{Rudolf Peierls Centre for Theoretical Physics, Oxford University, OX1 3NP, U.K.}

\begin{abstract}
Lattice Boltzmann simulations of liquid-gas systems are believed to be
restricted to modest density ratios of less than 10 \cite{Inamuro}. In
this article we show that reducing the speed of sound and, just as
importantly, the interfacial contributions to the pressure allows
lattice Boltzmann simulations to achieve high density ratios of 1000
or more. We also present explicit expressions for the limits of the parameter
region in which the method gives accurate results. There are two
separate limiting phenomena.  The first is the stability of the bulk
liquid phase. This consideration is specific to lattice Boltzmann
methods. The second is a general argument for the interface
discretization that applies to any diffuse interface method.
\end{abstract}

\maketitle

Simulations of liquid gas systems with lattice Boltzmann have been
restricted to small density ratios in the past. Those restrictions
have lead to the development of hybrid lattice Boltzmann schemes to be
able to simulate systems with high density ratios of about 100-1000 by
Inamuro \textit{et al.}\cite{Inamuro}. In this article we explain how
large density ratios can also be achieved with standard lattice
Boltzmann methods. Furthermore we derive the conditions which limit
the ability of the method to obtain stable, accurate, and unique
results for the phase diagram. We present a new general argument for
the minimum interface width required to accurately simulate a system
at a given reduced temperature. This important argument is not
restricted to lattice Boltzmann methods but only relies on the
relation of the discretized interface to the expression for the
pressure. It therefore applies to all diffuse interface methods.

The lattice Boltzmann method can be viewed as a discretization
of the Boltzmann equation. The hydrodynamic limit of the
Boltzmann equation gives the continuity and Navier-Stokes equations
and the discretization of the lattice version is chosen such that it
preserves this limit. The basic variables of the lattice Boltzmann
equation are a set of densities $f_i(\mb{x},t)$ associated with a
velocity set $\mb{v}_i$. The evolution equation for the $f_i$ is then
given by \cite{Me}
\begin{equation}
f_i(\mb{x}+\mb{v}_i,t+1)=
f_i(\mb{x},t)+F_i+\frac{1}{\tau}(f_i^0(\mb{x},t)+A_i-f_i(\mb{x},t)).
\label{LB}
\end{equation}
The $f_i^0$ are the equilibrium distribution corresponding to the
ideal gas. Non-ideal contributions are included through the bulk
forcing term $F_i$ or pressure term $A_i$, following
\cite{Me}. The fluid density is defined as $\rho=\sum_if_i$, and the
momentum is $\rho\mb{u}=\sum_if_iv_i$ (although the total momentum contains
additional contributions from the force).  The moments of the
equilibrium distribution are
\begin{eqnarray}
&\sum_if_i^0=\rho,\;\;\;\sum_if_i^0{v}_{i\alpha}=\rho{u}_\alpha,
\nonumber\\&
\sum_if_i^0{v}_{i\alpha}{v}_{i\beta}=\rho{u}_\alpha
  {u}_\beta +\rho\theta\delta_{\alpha\beta},
\nonumber\\
&\sum_if_i^0{v}_{i\alpha}{v_{i\beta}}{v}_{i\gamma}=\rho\theta (u_\alpha \delta_{\beta\gamma}+u_\beta
  \delta_{\alpha\gamma}+u_\gamma\delta_{\alpha\beta})
\nonumber\\&+\rho u_\alpha u_\beta u_\gamma+Q_{\alpha\beta\gamma}. 
\label{moment}
\end{eqnarray}
Here $Q$ is a correction term that should be zero. Most velocity sets
for lattice Boltzmann are limited to $v_{ix}\in\{-1,0,1\}$ so that
$v_{ix}^3=v_{ix}$. This restricts the third moment in (\ref{moment})
to $\theta=1/3$ and $Q_{\alpha\beta\gamma}=-\rho u_\alpha u_\beta u_\gamma$.

The non-ideal contributions from the $A_i$ need to conserve mass and
momentum and the moments are given by
\begin{eqnarray}
&\sum_i A_i = 0,\;\;\;
\sum_i A_i (\mb{v}_i-\mb{u}) = 0;\nonumber\\
&\sum_i A_i(v_{i\alpha}-u_\beta)(v_{i\beta}-u_\beta)
= A_{\alpha\beta},\\
&\sum_i A_i(v_{i\alpha}-u_\alpha)(v_{i\beta}-u_\beta)
(v_{i\gamma}-u_\gamma)
= A_{\alpha\beta}u_\gamma+A_{\alpha\gamma}u_\beta\nonumber\\&+A_{\beta\gamma}u_\alpha.\nonumber 
\end{eqnarray}
The forcing term $F_i$ has the moments
\begin{eqnarray}
&\sum_i F_i=0,\;\;\;
\sum_i F_i \left(\mb{v}_i-\mb{u}\right)=\mb{F},\nonumber\\
&\sum F_i \left(\mb{v}_i- \mb{u}\right) 
\left(\mb{v}_i- \mb{u}\right)
=\Psi.&
\end{eqnarray}

A standard expansion of (\ref{LB}) gives the continuity equation
\begin{equation}
\partial_t \rho+\nabla(\rho\h{u})=0
\end{equation}
and the Navier Stokes equation
\begin{equation}
\partial_t(\rho\h{u})+\nabla(\rho\h{u}\h{u})=
-\nabla(\rho\theta+A)+\mb{F}+\nabla\sigma+\nabla R
\label{NSeqn}
\end{equation}
where $\h{u}=\mb{u}+\frac{1}{2}\mb{F}$ \cite{Me}. Here the Newtonian stress tensor is
given by
\begin{equation}
\sigma=\nu \rho (\nabla\h{u}+(\nabla\h{u})^T)
\label{sigma}
\end{equation}
and unphysical terms have been collected in the remainder tensor
\begin{equation}
R=\tau \Psi  -3\nu [\h{u}\nabla.A+(\h{u}\nabla.A)^T+\h{u}.\nabla A \mb{1}+\nabla Q]+ O(\partial^2).
\label{eqnR}
\end{equation}
The kinematic viscosity is given by $\nu =(\tau-\frac{1}{2})\theta$.  Note that
most of the unphysical terms in (\ref{eqnR}) violate Galilean
invariance \cite{Holdych,Li}.

We see from (\ref{NSeqn}) that for $A=0$ and $F=0$ the lattice
Boltzmann method enforces an ideal gas equation of state with $p(\rho,\theta)=\rho\theta=\rho/3$.
To simulate a fluid with a non-ideal equation of state
$P(\rho,\theta)=\rho\theta+P^{nid}(\rho,\theta)$ we can now choose
\begin{eqnarray}
F &=& -\nabla.P^{nid},\label{forceeqn}\\
\Psi &=& (\tau-\frac{1}{4})\rho FF+\frac{1}{12}(\nabla\nabla)^D\rho,\nonumber\\
A &=& 0,\nonumber
\end{eqnarray}
which we will refer to as the forcing method \cite{Me}. The careful
reader may have recognized that the $\Psi$ term does not appear in the
Navier-Stokes equation (\ref{NSeqn}), but a higher order analysis
shows that these terms are necessary to recover the correct
equilibrium behavior \cite{Me}. An alternative choice for the moments
is 
\begin{eqnarray}
&&F = 0,\;\;\; \Psi = 0,\nonumber\\
&&A = P^{nid}+\nu (\h{u}\nabla\rho+(\h{u}\nabla\rho)^T+\h{u}.\nabla\rho \theta \mb{1}),
\label{pressure1}
\end{eqnarray}
which we will call the pressure method \cite{Holdych,Me}.
For either approach we recover the Navier Stokes equation for a
non-ideal gas
\begin{equation}
\partial_t(\rho\h{u})+\nabla(\rho\h{u}\h{u})=
-\nabla P+\nabla\sigma.
\end{equation}
In equilibrium both approaches lead to a constant pressure $P$ and
therefore to the same density profiles\cite{Me}.

\begin{figure}
\includegraphics[width=0.8\columnwidth]{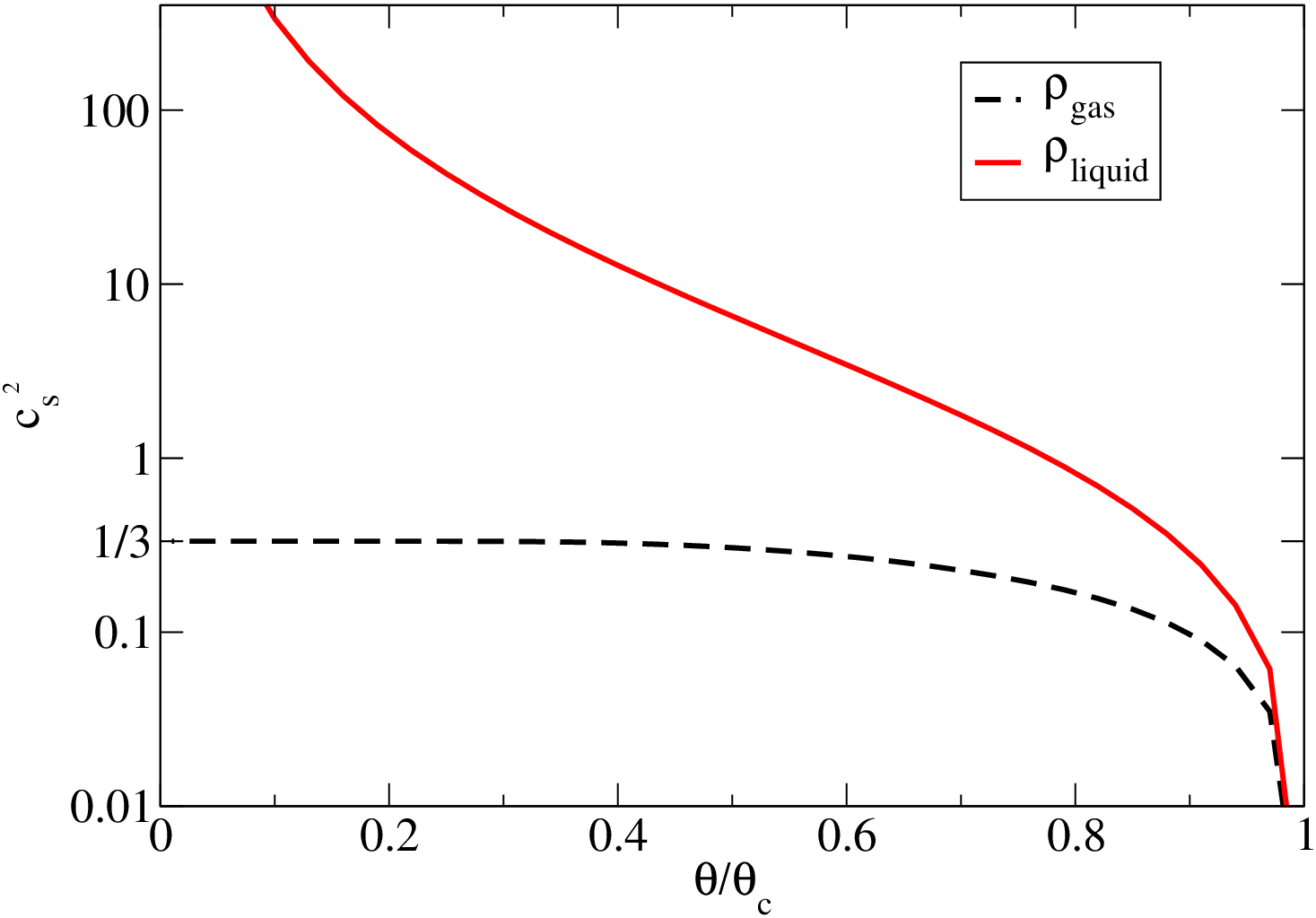}

\caption{Sound speed squared $c_s^2$ for the liquid and gas phases as
  a function of the quench depth for $p_0=1$.}
\label{fig_csT}
\end{figure}

Most previous lattice Boltzmann simulations approached the simulation
of non-ideal systems by using the ideal gas equation of state
$p=\rho\theta=\rho/3$, as a starting point.  Interactions are then included to
allow the simulation of non-ideal systems. The speed of sound
$c_s=\sqrt{\partial_\rho p}$ will then recover the ideal gas value of 1/3 in the
dilute limit. For a van der Waals gas with a critical density of 1 and
a temperature of $\theta=1/3$ and an interfacial free energy of $\int
\kappa/2\;(\nabla\rho)^2$ the pressure tensor is given by
\begin{eqnarray}
P&=&p_0\left[\left(\frac{\rho}{3-\rho}-\frac{9}{8}
  \rho^2\theta_c\right)\mb{1}
\right.\nonumber\\&&\left.
-\kappa\left(\rho\nabla^2\rho+\frac{1}{2}\nabla\rho.\nabla\rho\right)\mb{1}+\kappa\nabla\rho\nabla\rho
\right].
\label{VdW_int_new}
\end{eqnarray}
Previous approaches matched the ideal gas equation of state in the
dilute limit, leading to $p_0=1$.
For the van der Waals gas the speed of sound increases rapidly for
high densities. A problem arises when
the speed of sound becomes larger then the lattice velocity $|v_i|$,
because information can not be passed on at speeds larger than the
lattice velocity. When the speed of sound is increased above 1 the
simulation becomes unstable. This clearly limits the range of critical
temperatures for which we can obtain stable solutions in lattice
Boltzmann, as shown in Figure \ref{fig_csT}.

The stability analysis is slightly complicated by the fact that we
have additional gradient terms in the pressure tensor. These terms
further decrease the stability, as shown in a previous analysis of the
pressure method by C. Pooley for one, two, and three dimensional
lattice Boltzmann methods \cite{PooleyThesis}. In the notation of this
letter the linear stability condition is
\begin{equation}
c_s < \sqrt{1-4p_0\kappa \rho}
\label{stabeqn}
\end{equation}
for a homogeneous system with density $\rho$. This suggests that, at
least as far as the stability of the bulk phase is concerned, the most
stable solutions should be found for $\kappa=0$. 

\begin{figure}

\includegraphics[width=0.8\columnwidth]{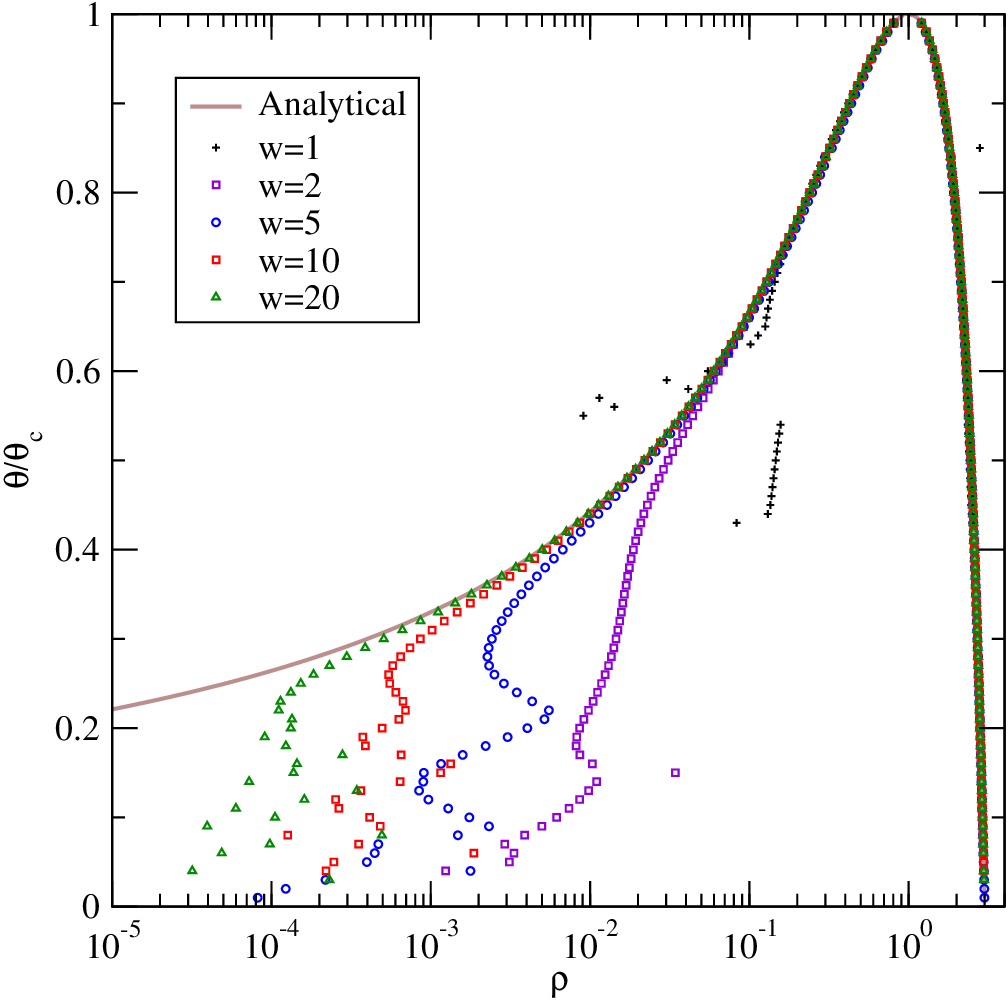}

\caption{The recovery of the phase diagram for different values of the
interface width. The van der Waals phase diagram is recovered to very
good approximation. For very deep quenches corresponding to large
density ratios wide interfaces are required to recover the very low
gas densities. The value of $p_0$ does not affect the form of the
interface or the value of the gas and liquid densities and values from
1 to $10^{-7}$ were used for increasing quench depth. }
\label{FigPhase}
\end{figure}

To lower the speed of sound in the liquid phase we now reduce the value
of $p_0$ in (\ref{VdW_int_new}). This decreases the speed of sound in
the liquid by a factor $p_0$. This also increases the range of
stability for $\kappa$ in (\ref{stabeqn}). We now expect that lowering the speed of
sound by a sufficient factor will reduce the speed of sound
sufficiently to simulate systems with arbitrarily low temperature
ratios $\theta/ \theta_c$. 

\begin{figure}
\subfigure[Pressure Method]{
\includegraphics[width=0.8\columnwidth]{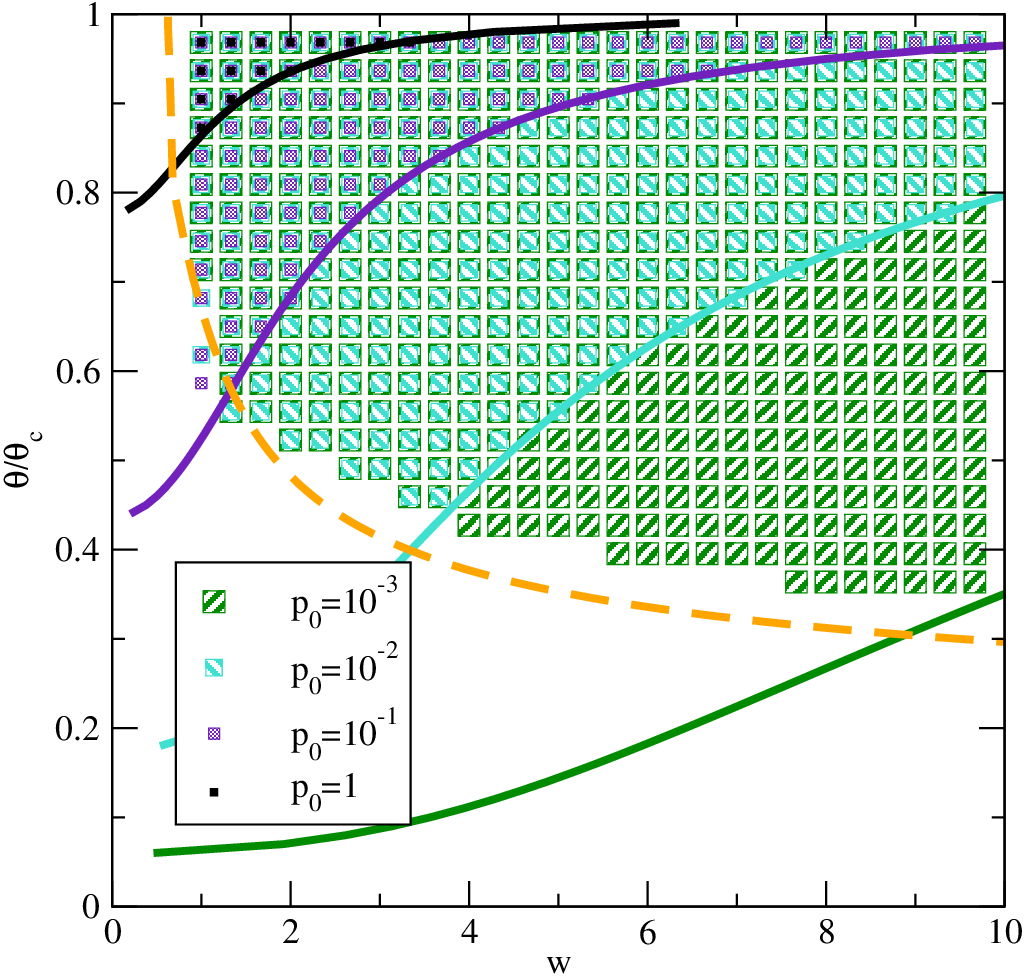}
}
\subfigure[Forcing Method]{
\includegraphics[width=0.8\columnwidth]{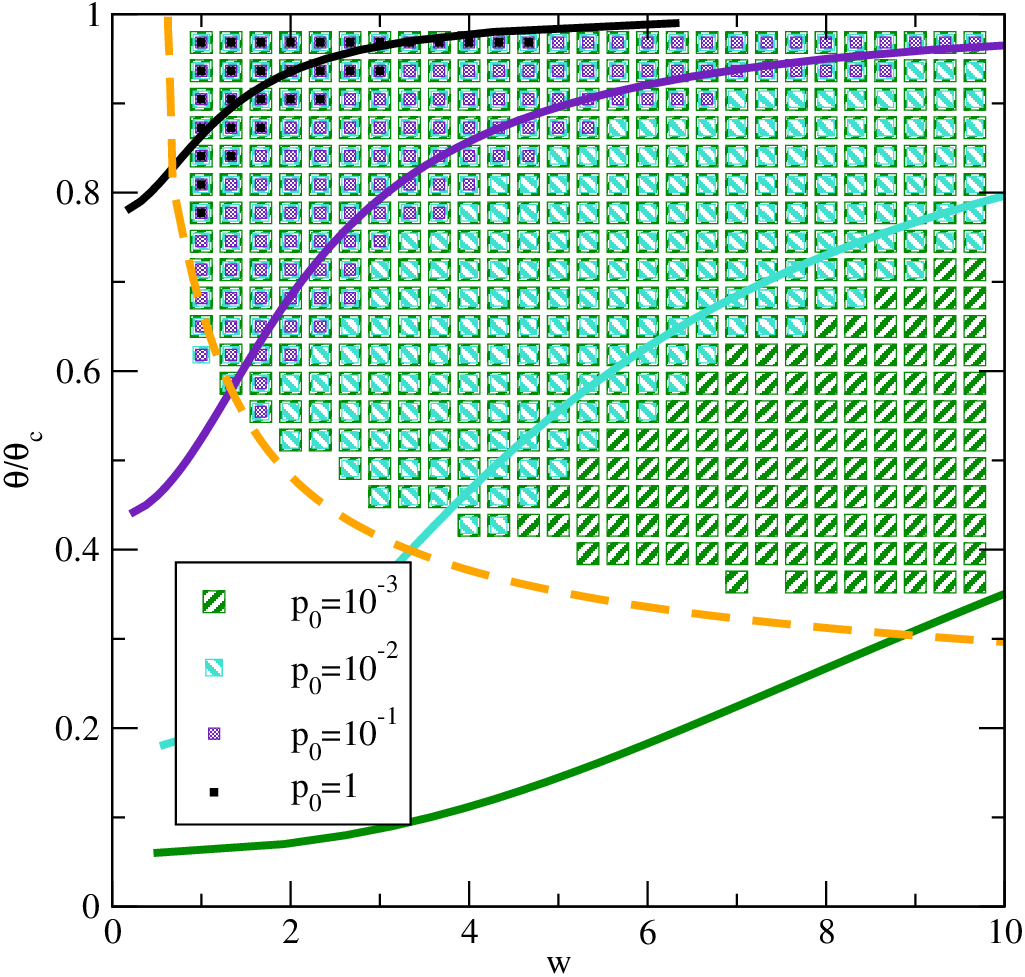}
}
\caption{Existence of accurate solutions for the (a) pressure and (b)
  corrected forcing method for different values of $p_0$ and
  $w$. Symbols indicate parameter combinations that lead to stable,
  accurate, and unique solutions. Solid lines are the bulk stability
  limits for the pressure method given by eqn. (\ref{stabeqn}). The
  dashed line is the line for an accurate interface representation
  given by eqn. (\ref{kappamin}).}
\label{Stab_1} 
\end{figure}

To test this idea we performed simulations with near equilibrium
profiles using a one dimensional three velocity $v_i=\{-1,0,1\}$ model
by defining
\begin{eqnarray}
\rho^{init}(x)&=&\rho_g+\frac{\rho_l-\rho_g}{2}(1+ 
\nonumber\\ 
&&\tanh\left[\frac{1}{w(\kappa,\theta/\theta_c)}\left(\left|x-\frac{N_x}{2}\right|-\frac{N_x}{4}\right)\right].
\label{profile}
\end{eqnarray}
where $\rho_l$ and $\rho_g$ are the equilibrium gas and liquid densities and
$N_x$ is the number of lattice sites. 
The interface width is given by
\begin{equation}
w(\kappa,\theta/\theta_c)=\sqrt{\frac{2\kappa}{\theta_c/\theta-1}}
\label{width}
\end{equation}
 This profile is not the exact
analytical solution to the differential equation $\nabla P=0$, but it is
very close to it. By initializing the simulation with this profile we
can test the linear stability of the method around an equilibrium
profile to good accuracy. The shape of a stable interfacial profile is
independent of $p_0$ for both the pressure and the forcing method.

In Figure \ref{FigPhase} we see that by lowering $p_0$ the method is
now able so simulate very small values of the reduced temperature $\theta/
\theta_c$ for interface width $w>1$, but that significantly
larger width are required to recover an accurate phase diagram for
deep quenches. For values of $\theta/ \theta_c$ between 0.9 and 1 we also find
non unique solutions for small values of $\kappa$ which is discussed in
more detail in a previous paper \cite{Me}.

To understand when the method fails to obtain accurate results note
that, in equilibrium, equation (\ref{VdW_int_new}) requires that
\begin{equation}
\kappa=\frac{\rho/(3-\rho)-9/8\;\rho^2\theta_c-p_b}{\rho\nabla^2\rho-\frac{1}{2}\nabla\rho.\nabla\rho}
\end{equation}
where $p_b$ is the bulk pressure corresponding to the density $\rho$.
For small values of $\kappa$ the interface becomes sharp in the continuous
limit so that the derivatives become arbitrarily large. But in the
numerical implementation the derivatives are {\em discrete}
derivatives. The discrete values are limited by the lattice
spacing. In the one dimensional case we choose $\nabla\rho(x) =0.5(\rho(x+1)-\rho(x-1))$
and $\nabla^2\rho(x)=\rho(x+1)-2\rho(x)+\rho(x-1)$. For higher dimensional stencils with
the same stability limits for the bulk phase see C. Pooley's thesis
\cite{PooleyThesis}.

The methods always lead to a constant pressure, even across an
interface\cite{Me}. We can now perform a simple estimate of the
minimum value $\kappa_{m}$ that allows this pressure to be the
equilibrium pressure. For any point with density $\rho_s$ we can consider
two neighboring points, one with a smaller density $\rho_-$ and one with
a larger density $\rho_+$. We can now find a lower limit for the smallest
value $\kappa_{m}$ by varying the values of $\rho_+$ and $\rho_-$
\begin{equation}
\kappa_{m}=\max_{\rho_g<\rho_s<\rho_l}\min_{
\begin{array}{c}
\scriptstyle \rho_l<\rho_-<\rho_s\\
\scriptstyle \rho_s<\rho_+<\rho_l
\end{array}}\frac{\rho/(3-\rho)-9/8\;\rho^2\theta_c-p_b}{\rho_s(\rho_--2\rho_s+\rho_+)-\frac{1}{8}(\rho_+ -\rho_-)^2}
\label{kappamin}
\end{equation}
where $\rho_l$ is the liquid density and $\rho_g$ is the gas density.  We
performed a scan of the parameter space $w$ and $\theta/ \theta_c$ initializing
the simulation with a near equilibrium profile for different values of
$p_0$. We will accept simulations that are stable, accurate and
unique. We choose as the criterion of accuracy that
$\log_{10}(\rho_{min})-\log_{10}(\rho_g)<0.1$. As can be seen in Figure
\ref{FigPhase}, the results are not very sensitive to the exact value
of the cutoff. For values of the interface width $w<1.5$ we also test
the uniqueness of the simulation by using initial profiles with bulk
densities corresponding to the pressure at the spinodal
points\cite{Me}. Our criterion for uniqueness is then that all
simulations lead to the same minimum density to within $\Delta\rho<0.01$.

The comparing (\ref{kappamin}), shown as a dashed line in Figure
\ref{Stab_1}, and the numerical results for stable, accurate and
unique solutions shows excellent agreement. The bulk stability of
eqn. (\ref{stabeqn}) gives the second limit for the acceptable
parameter range for the pressure method.  The forcing method leads to
a slightly larger range of bulk stability. The underlying reason is
that calculating derivatives in (\ref{forceeqn}) leads to an additional
information exchange allowing for speeds of sound slightly larger than
1. But the dependence of the stability on $p_0$ and $w$ is very
similar to the one for the pressure method.  Note that previous
lattice Boltzmann simulations correspond to $p_0=0$ which corresponds
to a small acceptable parameter range.

The interface constraint (\ref{kappamin}) is remarkably successful at
predicting the acceptable simulation parameters. It predicts how thin
is too thin for an interface. It thereby detects when non-unique
solutions occur and when solutions for deep quenches fail to deliver
accurate results. The criterion presented so far is entirely numerical
but because of its importance we want to examine two limiting cases
for shallow and deep quenches for which we can obtain analytical
results. 

\begin{figure}
\includegraphics[width=0.9\columnwidth]{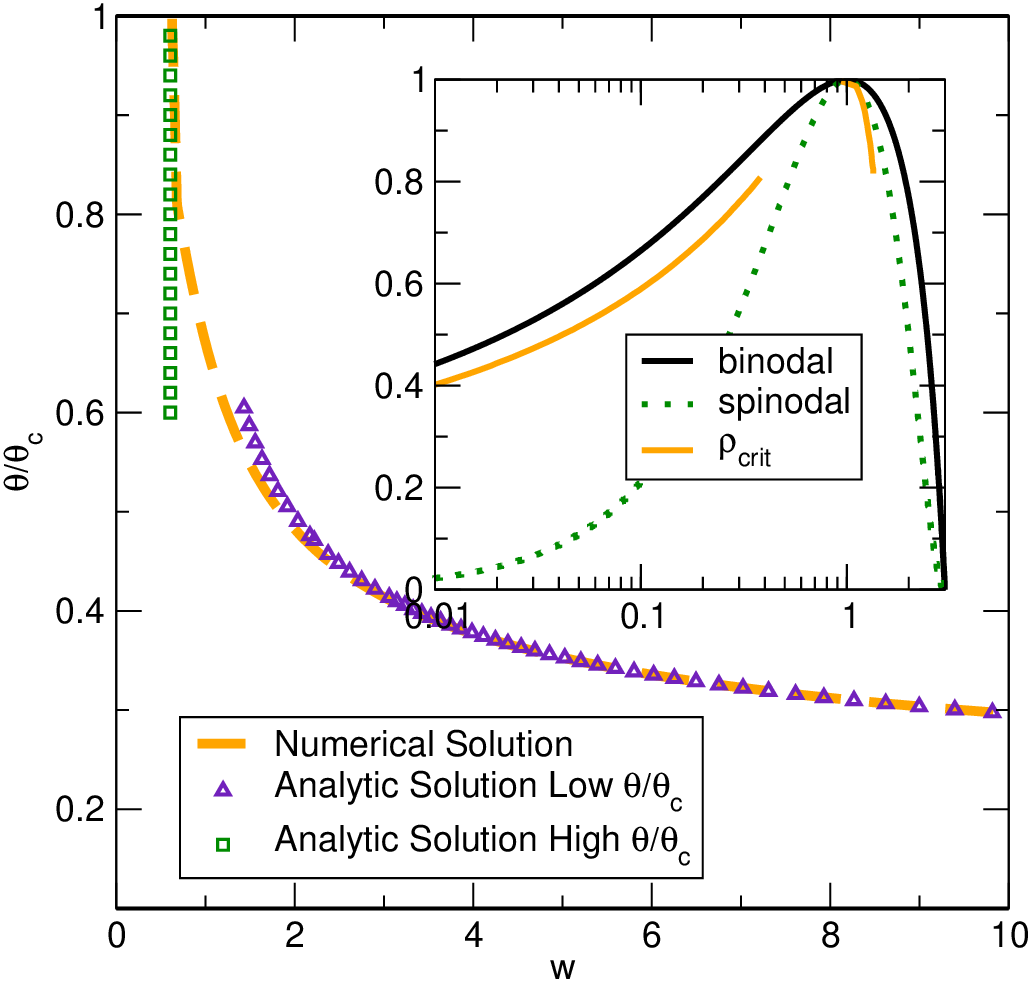}
\caption{The two limiting cases for which we can obtain an analytical
  approximation to the $w(\kappa_{m})$ relation. The inset shows the
  value of the critical density $\rho_{crit}$ for which the most severe
  limitation for $\kappa_{m}$ (\ref{kappamin}) occurs. Note that
  there is a discontinuity.}
\label{thetaw}

\end{figure}

We first examine for which values of $\rho_s$ the minimum value of
$\kappa$ is reached in eqn. (\ref{kappamin}).  The dashed line in
the inset of Figure \ref{thetaw} shows how this density $\rho_{crit}$ varies as a
function of temperature. 

Near to the critical temperature, the orange line in the inset in Figure
\ref{thetaw} lies close to the high density spinodal curve.  This is
because $P-p_b$ has its highest magnitude here, therefore helping to
maximize $\kappa_{m}$ within this region.  An analytical estimate for
$\kappa_{m}$ can be obtained by expanding the pressure around the
critical density, giving
\begin{eqnarray} 
P-p_b = -\frac{9}{4} \left( \theta_c-\theta \right) \left(\rho_s-1 \right) + \frac{3}{16} \left(\rho_s-1 \right)^3.
\end{eqnarray}
Within this regime, $P-p_b$ is large and negative, therefore $\rho_-$ and
$\rho_+$ must be chosen to make the denominator in (\ref{kappamin}) as
negative as possible.  A suitable choice is $\rho_-$ = $\rho_g$ and $\rho_+$ =
$\rho_s$.  We assume that the critical value of $\rho_s$ lies on the
spinodal curve $\rho_{spin} = 1 + 2 \sqrt{\theta_c-\theta}$.  This allows us to
obtain $\kappa_{m}(\rho_{crit})$, and substituting this expression into
(\ref{width}) gives a minimum interface width of
\begin{eqnarray}
w_{min} = \frac{1}{\sqrt{1+\sqrt{3}}}. 
\end{eqnarray}
As the temperature is decreased in the inset of Figure \ref{thetaw} the
critical density $\rho_{crit}$ makes a discontinuous jump to a regime in
which it lies close to the gas density, $\rho_g$.  The minimum interface
width can, in this case, be analytically obtained by expanding
densities around $\rho_g$. We define $\rho_s = \rho_g + \delta \rho$ and $\rho_+ = \rho_- + \Delta
\rho$.  Since $P-p_b$ is a positive quantity, a suitable choice for $\rho_-$
is $\rho_- = \rho_s$.  Substituting these expressions into equation
(\ref{kappamin}) gives
\begin{eqnarray}
\kappa_{m} = \frac{\theta \delta \rho}{(\rho_g + \delta \rho)\Delta \rho - \Delta \rho^2/8},
\label{kappamin2}
\end{eqnarray}
Minimizing this with respect to $\Delta\rho$ leads to $\rho_s =
\Delta\rho/4$. Re-substituting this result back into equation
(\ref{kappamin2}), and maximizing with respect to $\delta$, we finally
obtain $\rho_{crit} = 2 \rho_g$.  Using this we can calculate the minimum
interface width,
\begin{eqnarray}
w_{min} = \frac{1}{\sqrt{4 \rho_g \left(\theta_c - \theta \right)}},
\end{eqnarray}
as shown by the triangles in Figure \ref{thetaw}. This closely follows
the numerical result at low temperatures.

We have therefore shown how to simulate deep quenches with lattice
Boltzmann and we were able to predict which simulation parameters will
lead to accurate results.


\vspace{-0.6cm}
\begin{thebibliography}{99}
\bibitem{Inamuro} T. Inamuro, T. Ogata, S. Tajima, N. Konishi,
  \textit{J. Comp. Phys.}\textbf{198}, 628 (2004). 
\bibitem{Me} A.J. Wagner, submitted to \textit{Phys. Rev. E}, preprint
  available at cond-mat/0607087.
\bibitem{Holdych} D.J. Holdych, D. Rovas, J.G. Georgiadis,
  R.O. Buckius, \textit{Int. J. of Mod. Phys. C} \textbf{9}, 1393 (1998).
\bibitem{Li} A.J. Wagner and Q. Li,
  \textit{Physica A} \textbf{362}, 105 (2005).
\bibitem{PooleyThesis} Christopher M. Pooley, ``Mesoscopic modelling
  techniques for complex fluids'', thesis, Oxford University, 2003.

\end{thebibliography}
\end{document}